\documentclass[bibyear]{aa}  
\usepackage{graphicx}
\usepackage{txfonts}
\usepackage{natbib}
\begin{document} 

   \title{Persistent nuclear burning in Nova Sgr 2016 N.4 (= V5856 Sgr =
          ASASSN-16ma) six years past its outburst\thanks{Table~1 is only
          available in electronic form at the CDS via anonymous ftp to
          cdsarc.u-strasbg.fr (130.79.128.5) or via
          http://cdsweb.u-strasbg.fr/cgi-bin/qcat?J/A+A/}}
   \author{U. Munari
          \inst{1}
          \and
          N. Masetti
          \inst{2,3}
          \and
          F.~M. Walter
          \inst{4}
          \and
          R.~E. Williams
          \inst{5}
          \and
          F.-J. Hambsch
          \inst{6}
          \and
          A. Frigo
          \inst{6}
          \and
          P. Valisa
          \inst{6}
          }
   \institute{  INAF-Osservatorio Astronomico di Padova, 36012 Asiago (VI), Italy,\\
              \email{ulisse.munari@inaf.it}
         \and
                INAF-Osservatorio di Astrofisica e Scienza dello Spazio, via Gobetti 101, 40129 Bologna, Italy           
         \and
                Departamento de Ciencias F\'isicas, Universidad Andr\'es Bello, Fern\'andez Concha 700, Las Condes, Santiago, Chile
         \and
                Department of Physics and Astronomy, Stony Brook University, Stony Brook, NY 11794-3800, USA
         \and
                Space Telescope Science Institute, 3700 San Martin Drive, Baltimore, MD 21218, USA
         \and
                ANS Collaboration, c/o Astronomical Observatory, 36012 Asiago (VI), Italy
             }

   \titlerunning{Nova V5856 Sgr}   

   \date{Received September 15, 1996; accepted March 16, 1997}

  \abstract{
We report on the fast Nova Sgr 2016 N.4 being surprisingly trapped in a
long-lasting and bright plateau ($\Delta I$$\geq$10~mag above quiescence)
six years past the nova eruption.  Very few other novae experience a similar
occurrence.  We carried out an intensive observing campaign collecting daily
$B$$V$$R$$I$ photometry and monthly high-resolution optical spectroscopy,
and observed the nova in ultraviolet and X-rays with {\it Swift} at five
distinct epochs.  The bolometric luminosity radiated during the plateau is
$\sim$4200~L$_\odot$ (scaled to the distance of the Galactic Bulge),
corresponding to stable nuclear burning on a 0.6~M$_\odot$ white dwarf.  A
stable wind is blown off at full width at zero intensity (FWZI)$\sim$1600
km/s, with episodic reinforcement of a faster FWZI$\sim$3400 km/s mass loss,
probably oriented along the polar directions.  The collision of these winds
could power the emission detected in X-rays.  The burning shell has an outer
radius of $\sim$25~R$_\odot$ at which the effective temperature is
$\sim$7600~K, values similar to those of a F0 II/Ib bright giant.  The
$\Delta m$$<$1~mag variability displayed during the plateau is best
described as chaotic, with the irregular appearance of quasi-periodic
oscillations with a periodicity of 15-17 days.  A limited amount of dust
($\approx$3$\times$10$^{-11}$~M$_\odot$) continuously condenses at
$T_{dust}$$\sim$1200~K in the outflowing wind, radiating
$L_{dust}\sim$52~L$_\odot$.}

   \keywords{Stars: novae, cataclysmic variables --- Stars: winds, outflows}
   \maketitle

\section{Introduction}

ASASSN-16ma was discovered on 2016 October 25.02 UT by the All Sky Automated
Survey for Supernovae as a $V$$\sim$13.7 mag transient
\citep{2016ATel.9669....1S}.  It was soon classified by
\citet{2016ATel.9678....1L} as a FeII-type nova, with rather narrow emission
lines, unresolved with a full width at half maximum (FWHM) equal to the
instrumental PSF, and noted for the absence of P Cyg absorptions to Balmer
lines.  A month later, on 2016 November 23.1 UT, \citet{2016ATel.9849....1R}
obtained an optical--near-IR spectrum of ASASSN-16ma that confirmed the FeII
classification and the general absence of P Cyg absorptions and remarked on
the prevailing low expansion velocities (FWHM$\sim$1400 km\,s$^{-1}$) and
the presence of only low-excitation emission lines (no HeI visible). 
\citet{2017IAUC.9286....1N} reported ASASSN-16ma to be the fourth nova in
Sagittarius for the year 2016 and assigned it the variable star name V5856
Sgr, which we adopt in the rest of this paper.

Two weeks past discovery, $\gamma$-ray emission from the nova was observed
by Fermi-LAT \citep{2016ATel.9736....1L}: undetected until  2016 November 8, V5856
Sgr suddenly turned into a strong $\gamma$-ray source, remaining active
(although declining) for the following nine days
\citep{2016ATel.9771....1L}.  In a reanalysis of these Fermi-LAT data,
\citet{2022ApJ...924L..17L}  notes the probable presence at 4$\sigma$
significance of a 545 sec periodicity.

In a comparative analysis of nova optical light curves,
\citet{2017MNRAS.469.4341M} discuss how a second component appears to
develop in parallel with the detection of $\gamma$ rays, and note how in
V5856 Sgr such a second component outshined by $\sim$2 mag the main
component associated with the normal expanding ejecta.  The presence of an
additional component in the optical  light curve of V5856 Sgr related to the
emergence of $\gamma$ rays was also noted  by
\citet{2017NatAs...1..697L}.

At radio wavelengths the energetic events leading to $\gamma$-ray emission
did not reverberate much.  The radio observations of V5856 Sgr summarized by
\citet{2021ApJS..257...49C} show only standard thermal emission associated
with the expanding ejecta, peaking approximately three years past the
optical maximum, as is typical of many normal novae
\citep{1979AJ.....84.1619H}.  None of the features usually associated with
shocks and consequent synchrotron emission are visible in the radio data of
V5856 Sgr, which is characterized by no early peak and a low brightness
temperature, actually one of the lowest on record.  In this regard it is
worth noting that the radio monitoring of V5856 Sgr started early, when
Fermi-LAT was still recording strong $\gamma$-ray emission, but the nova
remained below the radio detection threshold for the first three months. 
\citet{2021ApJS..257...49C} remark on how V5856 Sgr appears underluminous at
radio wavelengths for the 2.5 kpc distance they adopt.  The discrepancy with
the other radio novae is lifted, however, if the larger 6.4$-$7.0 kpc
distance derived by \citet{2017MNRAS.469.4341M} is adopted instead.

\begin{table*}
\caption[]{BVRI photometry of V5856 Sgr in 2021 and 2022.  The long table is
available in its entirety in electronic form only; a small fraction is shown here
for guidance on its format and content.}
\begin{center}
\begin{tabular}{cccccc}

\noalign{\smallskip}
\hline
\hline
\noalign{\smallskip}

 HJD       & Date & $V$    & $B-V$ & $V-R$ & $R-I$ \\
(-2459000) & (UT) &      &   &   &  \\
\noalign{\smallskip}
\hline
\noalign{\smallskip}
 684.901 & 2022-04-15.401 & 12.481 ~$\pm$0.009 & 0.536 ~$\pm$0.007 & 0.408 ~$\pm$0.009 & 0.243 ~$\pm$0.020  \\
 685.901 & 2022-04-16.401 & 12.457 ~$\pm$0.009 & 0.570 ~$\pm$0.010 & 0.445 ~$\pm$0.009 & 0.235 ~$\pm$0.022  \\
 686.902 & 2022-04-17.402 & 12.483 ~$\pm$0.009 & 0.524 ~$\pm$0.009 & 0.436 ~$\pm$0.009 & 0.255 ~$\pm$0.024  \\
 687.902 & 2022-04-18.402 & 12.481 ~$\pm$0.008 & 0.525 ~$\pm$0.008 & 0.423 ~$\pm$0.011 & 0.249 ~$\pm$0.022  \\
\noalign{\smallskip}
\hline
\noalign{\smallskip}
\end{tabular}
\end{center}
\label{tab:BVRI}
\end{table*}

\begin{table}
\caption[]{Log of the spectroscopic observations of V5856 Sgr.}
\begin{center}
\begin{tabular}{@{~}c@{~~}c@{~~}c@{~}c@{~}c@{~~}r@{~}}

\noalign{\smallskip}
\hline
\hline
\noalign{\smallskip}

Date & UT & expt & airmass & spectr. & telescope \\
     &    & (sec)&         &         &           \\

\noalign{\smallskip}
\hline
\noalign{\smallskip}

2021-06-29 & 07:40 & 1200 & 1.31 &CHIRON& SMARTS 1.55m  \\
2021-07-10 & 22:24 & 1200 & 3.65 & B\&C & Asiago 1.22m  \\ 
2021-07-18 & 21:58 & 3600 & 3.68 &Echelle& Varese 0.84m \\ 
2021-07-19 & 21:38 & 3600 & 3.69 &Echelle& Varese 0.84m \\ 
2021-07-30 & 04:54 & 1000 & 1.15 &CHIRON& SMARTS 1.55m  \\
2021-08-08 & 21:08 & 2700 & 3.77 &Echelle& Varese 0.84m \\ 
2021-08-09 & 20:35 & 1800 & 3.67 & B\&C & Asiago 1.22m  \\ 
2021-08-09 & 21:07 & 3600 & 3.79 &Echelle& Varese 0.84m \\ 
2021-08-10 & 20:32 & 3600 & 3.68 &Echelle& Varese 0.84m \\ 
2021-08-23 & 03:29 & 1200 & 1.22 &CHIRON& SMARTS 1.55m  \\
2021-08-24 & 20:41 & 4500 & 4.04 &Echelle& Varese 0.84m \\ 
2022-03-16 & 09:37 & 2000 & 1.10 &CHIRON& SMARTS 1.55m  \\
2022-04-02 & 08:25 & 1000 & 1.11 &CHIRON& SMARTS 1.55m  \\
2022-04-25 & 06:24 & 1500 & 1.20 &CHIRON& SMARTS 1.55m  \\
2022-05-15 & 06:03 & 1500 & 1.08 &CHIRON& SMARTS 1.55m  \\
2022-06-08 & 03:16 & 3000 & 1.04 &CHIRON& SMARTS 1.55m  \\

\noalign{\smallskip}
\hline
\noalign{\smallskip}
\end{tabular}
\end{center}
\label{tab:log}
\end{table}

\begin{table*}
\caption[]{Log of the {\it Swift}/UVOT observations of V5856 Sgr. Flux densities are in 
units of 10$^{-15}$ erg cm$^{-2}$ s$^{-1}$ \AA$^{-1}$.}
\begin{center}
\begin{tabular}{ccccc}

\noalign{\smallskip}
\hline
\hline
\noalign{\smallskip}

 Date    & start     & $UVW1$ [2600 \AA]  & $UVM2$ [2246 \AA]  &  $UVW2$ [1928 \AA]\\
         & time (UT) & (mag \& flux)    & (mag \& flux)    &  (mag \& flux)  \\

\noalign{\smallskip}
\hline
\noalign{\smallskip}

 2021 Aug. 28 & 02:28     & 13.36$\pm$0.04 & 14.12$\pm$0.04 & 13.87$\pm$0.04 \\
              &           & 18.0$\pm$0.7   & 10.5$\pm$0.3   & 15.1$\pm$0.5   \\

\noalign{\smallskip}

 2021 Sep. 18 & 06:02     & 13.14$\pm$0.04 & 13.91$\pm$0.04 & 13.67$\pm$0.04 \\
              &           & 22.2$\pm$0.9   & 12.7$\pm$0.3   & 18.2$\pm$0.6   \\

\noalign{\smallskip}

 2021 Oct. 15 & 04:50     & 12.64$\pm$0.04 & 13.28$\pm$0.04 & 13.14$\pm$0.04 \\
              &           & 35.1$\pm$1.3   & 22.6$\pm$0.5   & 29.6$\pm$0.8   \\

\noalign{\smallskip}

 2021 Nov. 05 & 13:52     & 13.00$\pm$0.04 & 13.58$\pm$0.04 & 13.43$\pm$0.04 \\
              &           & 25.2$\pm$1.0   & 17.1$\pm$0.5   & 22.7$\pm$0.7   \\

\noalign{\smallskip}

 2022 Apr. 27 & 11:39     & 13.25$\pm$0.04 & 13.97$\pm$0.04 & 13.79$\pm$0.04 \\
              &           & 19.9$\pm$0.07  & 12.0$\pm$0.3   &16.3$\pm$0.6   \\

\noalign{\smallskip}
\hline
\noalign{\smallskip}
\end{tabular}
\end{center}
\label{tab:Swift}
\end{table*}

The Swift satellite looked for X-ray emission from V5856 Sgr during the main
outburst, but none was observed.  Of the 13 novae emitting in $\gamma$ rays
and studied by \citet{2021ApJ...910..134G}, only two were not detected as
X-ray sources with Swift, namely V1324 Sco and V5856 Sgr.  The excessive
distance (6.5 kpc) was blamed for the nondetection of V1324 Sco, but the 2.5
kpc distance adopted by \citet{2021ApJ...910..134G} for V5856 Sgr made its
nondetection puzzling.  As for the radio luminosity above, also in this case
the 6.4$-$7.0 kpc distance derived by \citet{2017MNRAS.469.4341M} would
justify the nondetection in X-rays.  V5856 Sgr is not included in the latest
Gaia DR3 data release because it is based on observations collected by the
spacecraft prior to 28 May 2017, which is only a few months past the
eruption and is too early for any astrometric characterization.

In this paper we focus on the recent and unexpected behavior displayed by
V5856 Sgr, after we noted \citep{2021ATel14884....1M,
2021ATel14804....1M}  that six years past its outburst the nova
remains halfway to quiescence (still $\Delta I$$\geq$10 mag brighter than
that).  We   carried out daily $B$$V$$R$$I$ photometry, obtained
high-resolution spectroscopy at monthly cadence, and observed on multiple
epochs V5856 Sgr in X-rays and ultraviolet with the {\it Swift} satellite. 
A detailed analysis of the whole body of spectroscopic data collected on
V5856 Sgr during its entire evolution, including the main outburst, will be
the subject of a separate paper (R.  Williams et al., in prep.).

\section{Observations}

\subsection{Optical photometry}

Optical photometry of V5856 Sgr was obtained simultaneously in the
$B$$V$$R$$I$ bands during 108 nights in 2021 and 111 nights in 2022, with
the same robotic 40cm telescope (located in San Pedro de Atacama, Chile) and
the same observing procedures adopted by \citet{2017MNRAS.469.4341M} to
cover the main outburst, in particular ($i$) the same local photometric
sequence around V5856 Sgr was used to solve the color equations for each
observing night and accurately place the observations on the
\citet{2009AJ....137.4186L} photometric system, and ($ii$) photometry was
carried out in PSF-fitting mode on the central server of ANS Collaboration
in Asiago.  The results are given in Table~\ref{tab:BVRI}, where the quoted
uncertainty is the total error budget, adding quadratically all the sources
including the Poissonian noise on the variable and the error in the
transformation from the local instantaneous photometric system to the
Landolt {\bf equatorial standard system}.

\subsection{Optical spectroscopy}

Optical spectra of V5856 Sgr were recorded   from Italy and from Chile.  
Table~\ref{tab:log} provides a log-book for them.

From Italy, low-resolution spectroscopy of V5856 Sgr was obtained with the
Asiago 1.22m telescope equipped with a Boller and Chivens   (B\&C) spectrograph.  The CCD camera is an ANDOR iDus
DU440A with a back-illuminated E2V 42-10 sensor, 2048$\times$512 array of
13.5 $\mu$m pixels.  The long-slit spectra were recorded with a 300 ln/mm
grating blazed at 5000 \AA, and covered the wavelength range from 3300 to
8000 \AA\ at 2.31 \AA/pix.  The 2 arcsec slit was imaged at a FWHM(PSF)=2.5
pixel scale.  Echelle spectra of V5856 Sgr were obtained with the Varese
0.84m telescope, equipped with a mark.III Multi-Mode Spectrograph from
Astrolight Instruments.  The camera is a SBIG ST10XME CCD and the 4250-8850
\AA\ range is covered in 32 orders without inter-order gaps.  A 2x2 binning
and the slit widened to 3 arcsec reduced the resolving power to 11,000.  The
spectra from both Asiago and Varese were exposed with the slit rotated to
the parallactic angle, and the data reduced with IRAF\footnote{IRAF is distributed by
the National Optical Astronomy Observatories, which are operated by the
Association of Universities for Research in Astronomy, Inc., under
cooperative agreement with the National Science Foundation.}.  For both sites the
nova culminates at just 16$^\circ$ above the local horizon, imposing a large
airmass and thus  poor seeing.

We also observed V5856 Sgr from Chile, where it transits nearly overhead,
using the CHIRON \citep{2013PASP..125.1336T} fiber-fed bench-mounted Echelle
fed by the CTIO 1.5m telescope operated by SMARTS.  We used CHIRON in
``fiber'' mode with 4$\times$4 on-chip binning yielding a resolution
$\lambda/\delta\lambda \approx$27,800.  Exposure times range from 15 to 50
minutes, typically in co-added 15-20 minute integrations.  The eight spectra
obtained during the 2021 and 2022 observing seasons are listed in Table 2;
those obtained at earlier epochs will be discussed in R. Williams et al. (in prep.).

The data were reduced using a pipeline coded in  Interactive Data
Language (IDL).\footnote{http://www.astro.sunysb.edu/fwalter/SMARTS/CHIRON/ch\_reduce.pdf}
The images were flat-fielded.  Cosmic rays were removed using the L.A. 
Cosmic algorithm \citep{2001PASP..113.1420V}.  The 74 echelle orders were
extracted using a boxcar extraction, and the instrumental background  computed
on both sides of the spectral trace was subtracted.  As CHIRON is
fiber-fed there is no simple method to subtract the sky.  The fibers have a
diameter of 2.7~arcsec on the sky.  In any event, for bright targets the  night
sky emission is generally negligible, apart from narrow [OI] and NaI~D lines
and some OH airglow lines at longer wavelengths.  Wavelength calibration
uses ThAr calibration lamp exposures at the start and end of the night.

The instrumental response is removed from the individual orders by dividing
by the spectra of a flux-standard star, $\mu$~Col.  This provides
flux-calibrated orders with a systemic uncertainty due to sky conditions. 
Individual orders are spliced together, resulting in a calibrated spectrum
from 4083-8900 \AA, with five inter-order gaps in the coverage longward of
8260~\AA.  Contemporaneous $B$$V$$R$$I$ photometry from Table~1 was
used to scale the spectra to approximately true fluxes.

\subsection{Swift UVOT and XRT}

A series of higher-energy observations of V5856 Sgr were acquired with the
{\it Swift} satellite \citep{2004ApJ...611.1005G}.  The pointings were
carried out in target-of-opportunity mode; these  observations are
generally limited to roughly 2000~s per visit, and four of them were
performed on the source with nearly monthly cadence between August and
November 2021.  A final fifth   observation was acquired in April 2022.  The dates and
start times are reported in Table~\ref{tab:Swift}.

The {\it Swift} observations were acquired with the onboard instruments
X-Ray Telescope \citep[XRT;][]{2005SSRv..120..165B} and UltraViolet Optical
Telescope \citep[UVOT;][]{2005SSRv..120...95R}.  The XRT allows  coverage
of the X-ray band between 0.3 and 10 keV, whereas UVOT data were collected
using the UV filters $UVW1$, $UVM2$, and $UVW2$, with reference wavelengths
2600 \AA, 2246 \AA, and 1928 \AA, respectively \citep[see][for
details]{2008MNRAS.383..627P,2011AIPC.1358..373B}.  On-source pointings were
simultaneously performed with the two instruments and lasted between
$\sim$1000 and $\sim$1800 s for XRT, whereas exposures between 101 and 629 s
were used for the three UVOT filters, depending on the observation.  All
data were reduced within the {\sc ftools} environment \citep{1995ASPC...77..367B}.

   \begin{figure}
   \centering
   \includegraphics[width=\hsize]{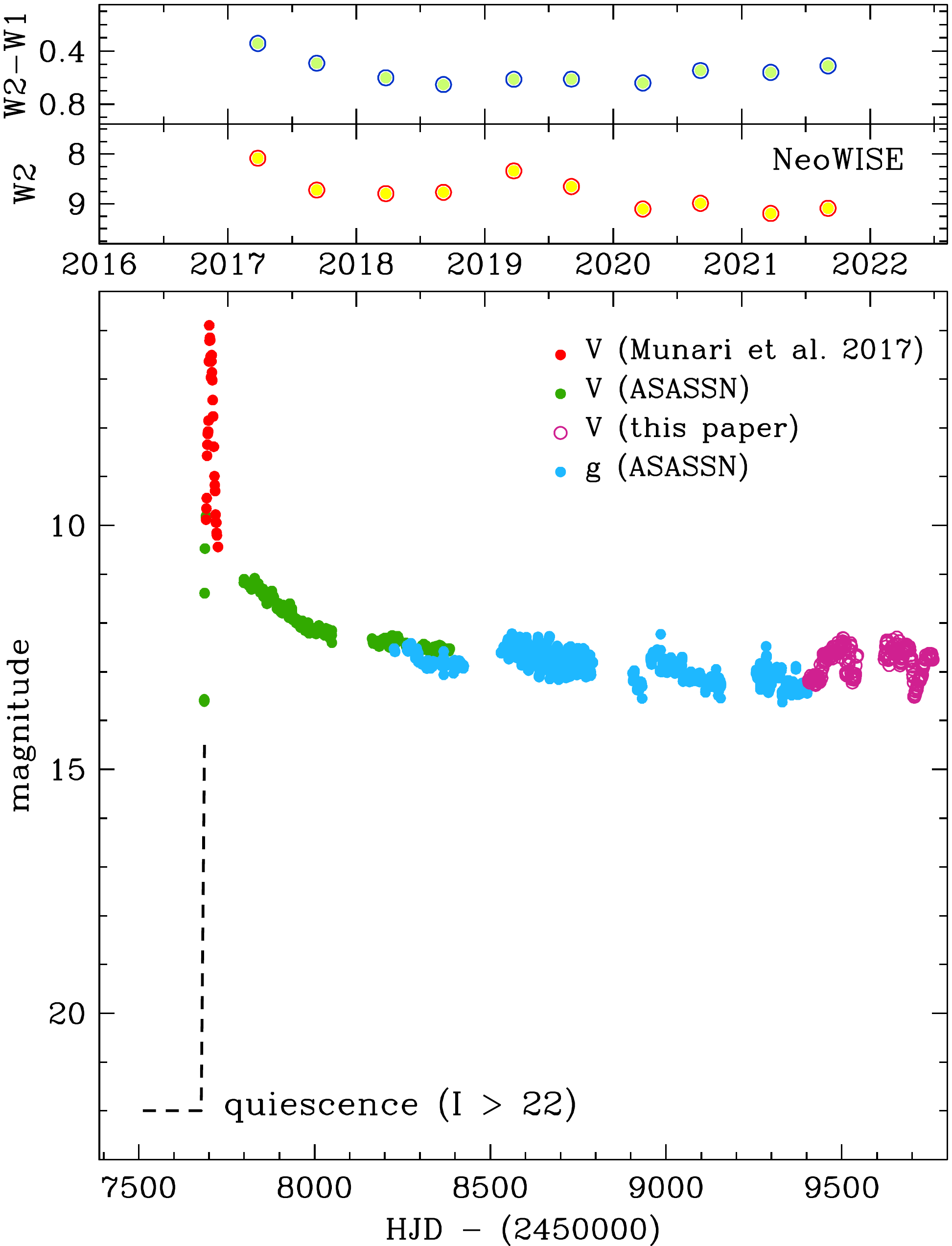}
      \caption{Complete optical light curve of V5856
              Sgr since its nova eruption in 2016. Top panel: 
              Infrared magnitudes from NeoWISE all-sky survey.}
         \label{fig:Overall}
   \end{figure}

   \begin{figure*}
   \centering
   \includegraphics[width=\hsize]{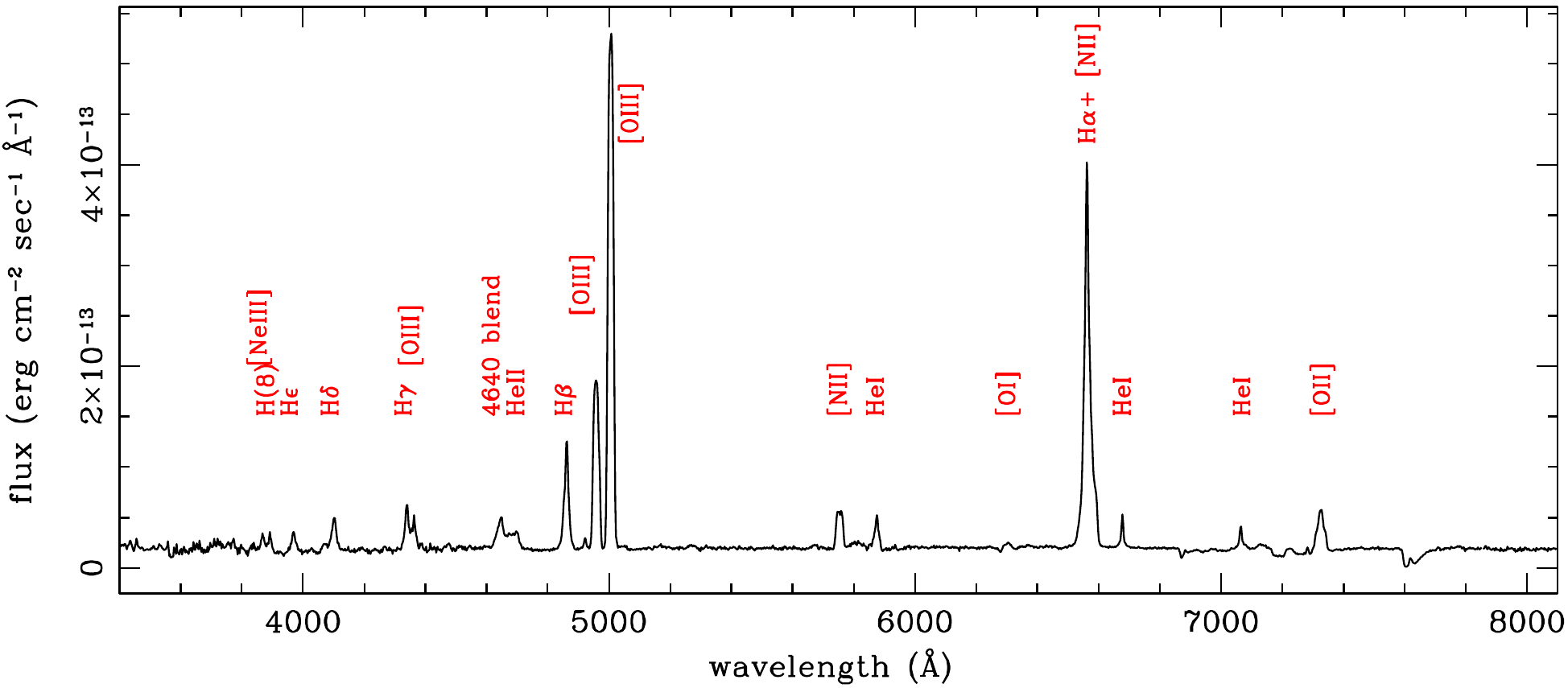}
      \caption{Spectrum of V5856 Sgr obtained on 2021 August 9 with the Asiago
               1.22m + B\&C telescope. The  strongest emission lines are identified.}
         \label{fig:122m}
   \end{figure*}

Count rates on Level 2 (i.e.,  calibrated and containing astrometric
information) UVOT images of V5856 Sgr were measured through aperture
photometry within a 5$''$ radius centered on the source position, whereas
the corresponding background was evaluated for each image using a
combination of several circular regions in source-free nearby areas.  The UV
magnitudes of V5856 Sgr were determined with the {\sc uvotsource} task.  The
data were then calibrated using the UVOT photometric system described by
\citet{2008MNRAS.383..627P}; the most recent fixings (2020 November) 
recommended by the UVOT team were taken into account.  The results of this
analysis are listed in Table~\ref{tab:Swift}.

The XRT data analysis was performed using the {\sc xrtdas} standard pipeline
package ({\sc xrtpipeline} v.  0.13.4) in order to produce screened event
files.  All X-ray data were acquired in photon counting (PC) mode 
\citep{2004SPIE.5165..217H}
adopting the standard grade filtering (0–12 for PC) according to
the XRT nomenclature.  Scientific data for V5856 Sgr were extracted from the
images using a radius of 47$''$ (20 pixels) centered again at the optical
coordinates of the source, while the corresponding background was evaluated
in a source-free region of radius 94$''$ (40 pixels) within the same XRT
acquisition.  In each single case no emission was detected in the 0.3–10
keV range using the {\sc xspec} package down to count rates between
$\sim$5$\times$10$^{-3}$ and $\sim$8$\times$10$^{-3}$ counts s$^{-1}$
(3$\sigma$ limits).  Nevertheless, by summing up the five XRT pointings, we
could reach a 5$\sigma$ detection of V5856 Sgr in the 0.3-10 keV band at a
rate of  (4.7$\pm$0.9)$\times$10$^{-3}$ counts s$^{-1}$, with the bulk of the
emission ($\sim$80\% of the counts) concentrated in the 0.3-2 keV range. 
Due to the low overall signal-to-noise ratio of the summed XRT observation, no
further detailed spectral analysis was performed.

We then determined the corresponding X-ray flux using the {\sc webpimms}
online tool\footnote{{\tt
https://heasarc.gsfc.nasa.gov/cgi-bin/Tools/ /w3pimms/w3pimms.pl}} by assuming
a thermal bremsstrahlung emission with temperature $kT$ = 1 keV plus an
intervening hydrogen column density absorption $N_{\rm H}$ =
1.8$\times$10$^{21}$ cm$^{-2}$ (obtained adopting the interstellar reddening
$E_{B-V}$=0.32 derived in Sect. 3 combined with the empirical formula of
\citet{1995A&A...293..889P}. This implies a count rate-to-flux conversion
factor of 2.4$\times$10$^{-11}$ erg cm$^{-2}$ s$^{-1}$ counts$^{-1}$; the
count rate reported above thus corresponds to absorbed and unabsorbed fluxes
of (1.1$\pm$0.2)$\times$10$^{-13}$ erg cm$^{-2}$ s$^{-1}$ and
(1.9$\pm$0.4)$\times$10$^{-13}$ erg cm$^{-2}$ s$^{-1}$, respectively, for
the assumed spectral model.

\section{Post-outburst evolution} 

   \begin{figure}
   \centering
   \includegraphics[width=\hsize]{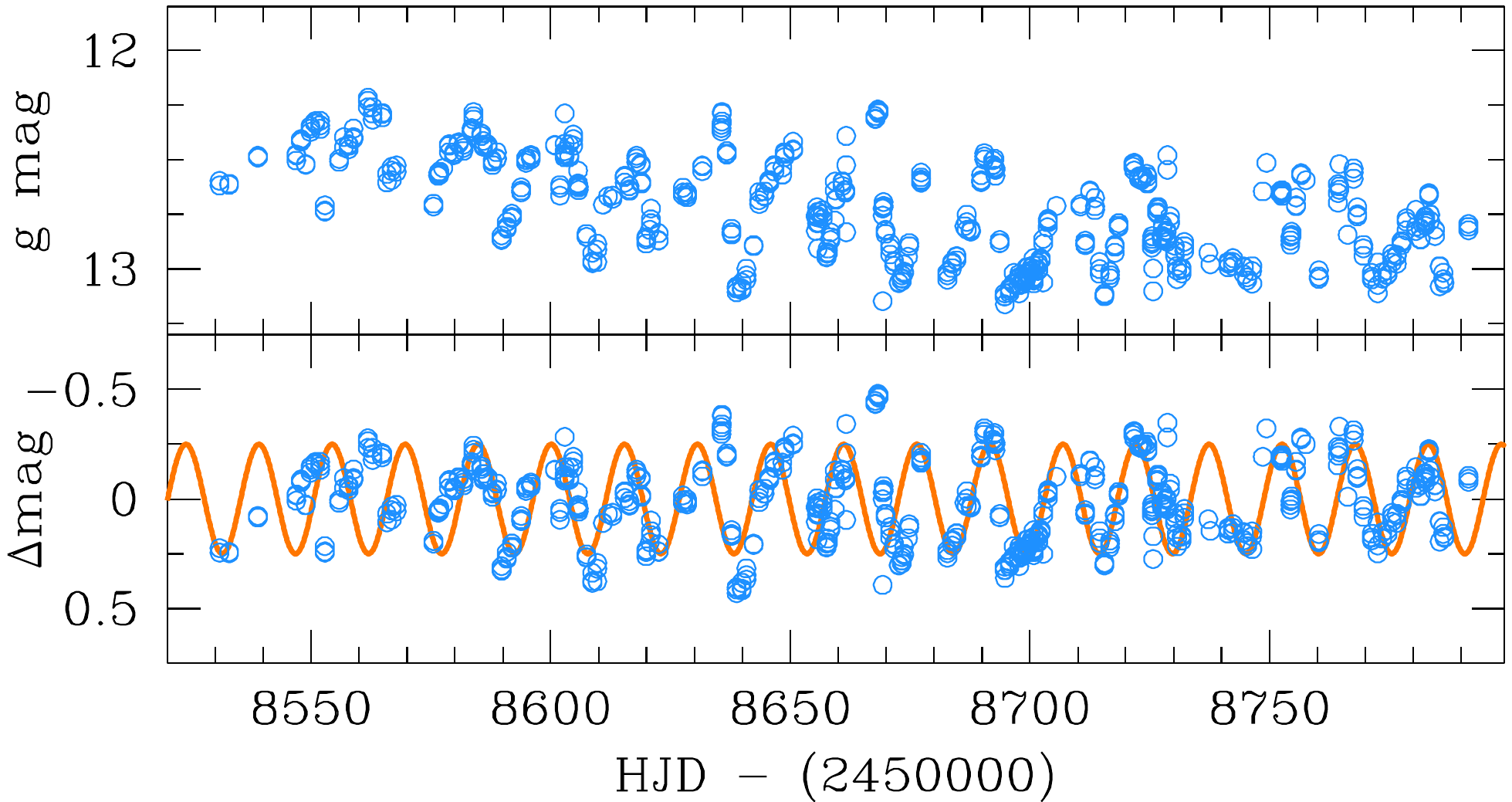}
      \caption{Portion of  2019  light curve of V5856 Sgr from
        Figure~\ref{fig:Overall}.  Top panel: Original ASASSN data in
        the $g$ band.  Bottom panel: Same data after detrending and with
        a superimposed  sinusoid of 15-day period and 0.25mag amplitude.}
         \label{fig:oscillations}
   \end{figure}

\subsection{A halted decline and the 2018-2022 plateau}

The overall light curve of V5856 Sgr since its outburst in late 2016 is
presented in Figure~\ref{fig:Overall}.  It was prepared with data from
\citet{2017MNRAS.469.4341M} for the main 2016 outburst, from this paper for
2021 and 2022, and from ASASSN for the interval in between
\citep{2014ApJ...788...48S,2017PASP..129j4502K}.  The quiescence level is
taken from \citet{2016ATel.9683....1M}, who inspected the OGLE-IV deep
template images and set an upper limit of $I$$>$22~mag to the brightness of
the progenitor in quiescence.

   \begin{figure*}
   \centering
   \includegraphics[angle=270,width=17cm]{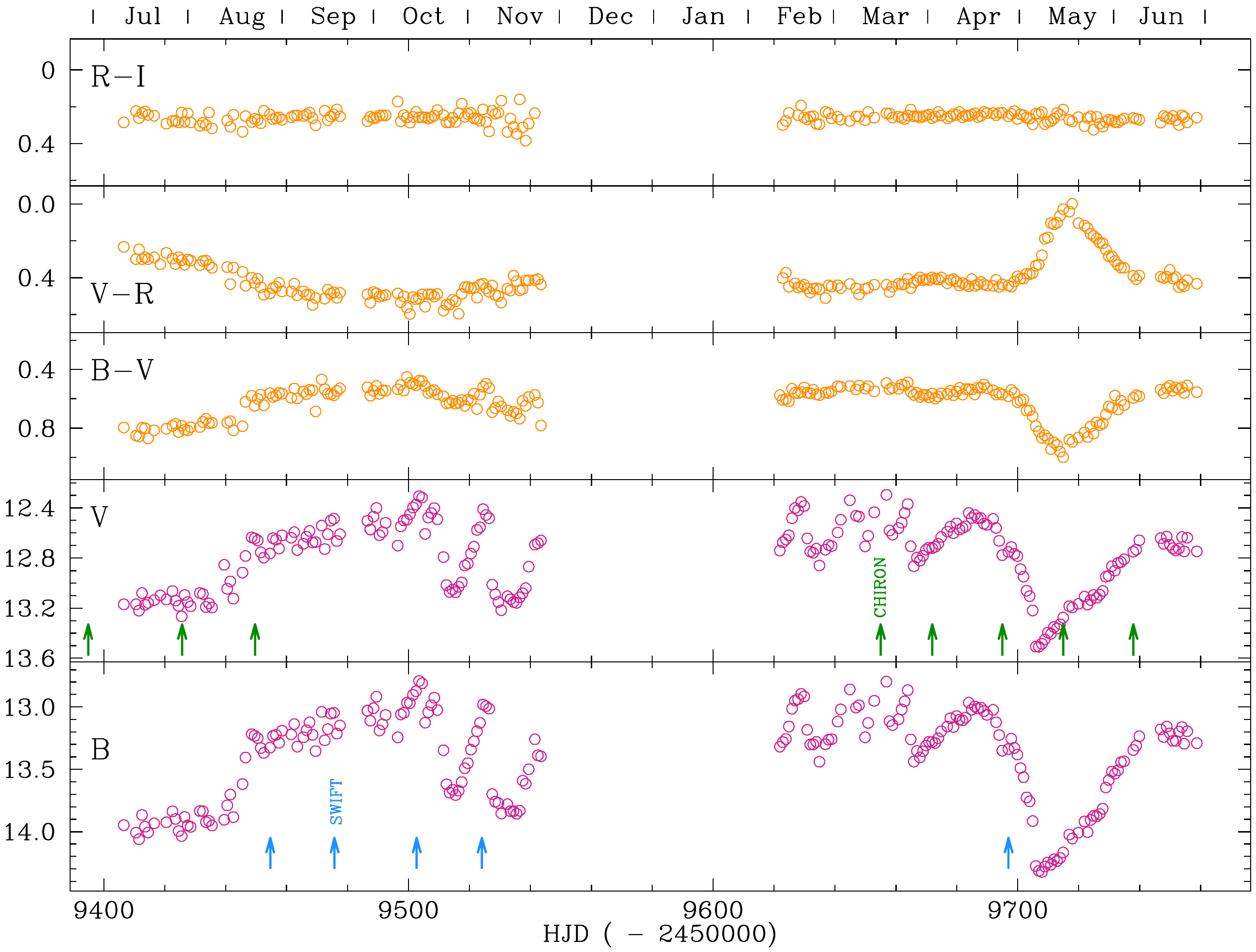}
      \caption{Light curves and color curves of V5856 Sgr in 2021-22 from  Table~\ref{tab:BVRI}.  The epochs of Swift and CHIRON
      observations are flagged  by arrows in the bottom panels.}
         \label{fig:BVRI}
   \end{figure*}

After a slower-than-expected early decline in 2017, since 2018 the nova has
not declined further, levelling out on a plateau, at median brightness
$I$=12.05, which is $\Delta I$$\geq$10 mag above the quiescence level.  A
halted decline is rather unusual among novae, and it is generally attributed
to protracted nuclear burning on the surface of the WD, as for V723 Cas =
Nova Cas 1995 \citep[e.g.,][]{2015MNRAS.454..123O, 2015ATel.7985....1G,
2015ATel.8053....1N, 2018AJ....155...58H}.  In their morphological grouping
of nova light curves, \citet{2010AJ....140...34S} defined a heterogeneous
$P$ class as composed of objects showing a temporary flattening of their
declines, lasting 15-500 days (median value $\sim$70 day; interestingly, the
group does not include V723 Cas).  While probably related to some members of
the $P$ class, the timescale of the V5856~Sgr plateau is at least one order
of magnitude longer.

The photometric stability of V5856 Sgr during the plateau extends to the
near-IR as well.  In Figure~\ref{fig:Overall} we  plot the results
gathered by the NeoWISE mission during its all-sky scanning, which  revisits
the position of V5856 Sgr twice a year (March and September).  NeoWISE
\citep{2011ApJ...731...53M, 2014ApJ...792...30M} refers to the data the WISE
satellite is collecting in the  W1 (3.4 $\mu$m) and W2 (4.6 $\mu$m) bands since
it was brought out of hibernation and resumed observation in 2014,
after the conclusion of the 2009-2010 cryogenic phase that also observed  in the 
W3 (12 $\mu$m) and W4 (22 $\mu$m) bands \citep{2014yCat.2328....0C}.  During
the 2018-2022 plateau, V5856 Sgr fluctuated by $\sigma$(W1)=0.25 mag around
the mean value $<$W1$>$=9.36 mag, and by $\sigma$(W1-W2)=0.06 mag around the
mean color $<$W1-W2$>$=+0.59 mag.

The spectral appearance of V5856 Sgr during the plateau is shown in
Figure~\ref{fig:122m}.  Prominent Balmer and HeI emission lines are
superimposed on a strong and featureless continuum, in particular around
the expected position for any Balmer discontinuity.  Intense [OIII], [NII],
and [OII] nebular lines are present, the corresponding strong auroral
transitions suggesting high electron densities, too high for the original
nova ejecta after several years of undisturbed expansion.  All lines are
resolved at a FWHM$\sim$1000 km/s, but with differences from line to line
(see Sect.  6 below).  The ionization degree is relatively low, with just a
weak HeII 4686 visible in emission; the criteria outlined by
\citet{1994A&A...282..586M} suggest $T_{\rm eff}$=5$\times$10$^4$~K for the
temperature of the photoionizing source.

Superimposed on a stable mean brightness during the plateau, V5856 Sgr 
presents some variability of limited amplitude, rather erratic in nature,
with the exception of 2019 when the nova displayed a persistent oscillation
superimposed on a mildly declining pattern, as illustrated in
Figure~\ref{fig:oscillations}.  We {\bf have carried out a} Fourier analysis of the
data in Figure~\ref{fig:oscillations}, which returned a low-significance
periodicity of $\sim$15 days.  In the lower panel of
  Figure~\ref{fig:oscillations} we  overplot on the observations a
sinusoid with a 15-day period and 0.25 mag semi-amplitude; the sinusoid
is followed by the data rather closely for only a few cycles at a time, and
then the correlation is lost.

\subsection{Multicolor evolution in 2021-2022}

After we called attention to its halted decline \citep{2021ATel14884....1M,
2021ATel14804....1M}, we started a daily $B$$V$$R$$I$ monitoring of V5856
Sgr.  The resulting light curves and color curves are presented in
Figure~\ref{fig:BVRI}, where the gap from late 2021 November to early 2022
February corresponds to the solar conjunction.

The photometric behavior of V5856 Sgr in Figure~\ref{fig:BVRI} is quite
erratic: flat and smooth in 2021 July and September, separated by a sudden
jump in August; the emergence of pronounced oscillations in 2021 October and
November, which continued through 2022 February, March, and April.  The
timescale of the oscillations observed in 2021-22 is $\sim$17 days, similar
to the 15-day pseudo-periodicity observed in 2019, where the brightness
levels of the nova at both epochs are similar.

   \begin{figure*}
   \centering
   \includegraphics[angle=270,width=17.8cm]{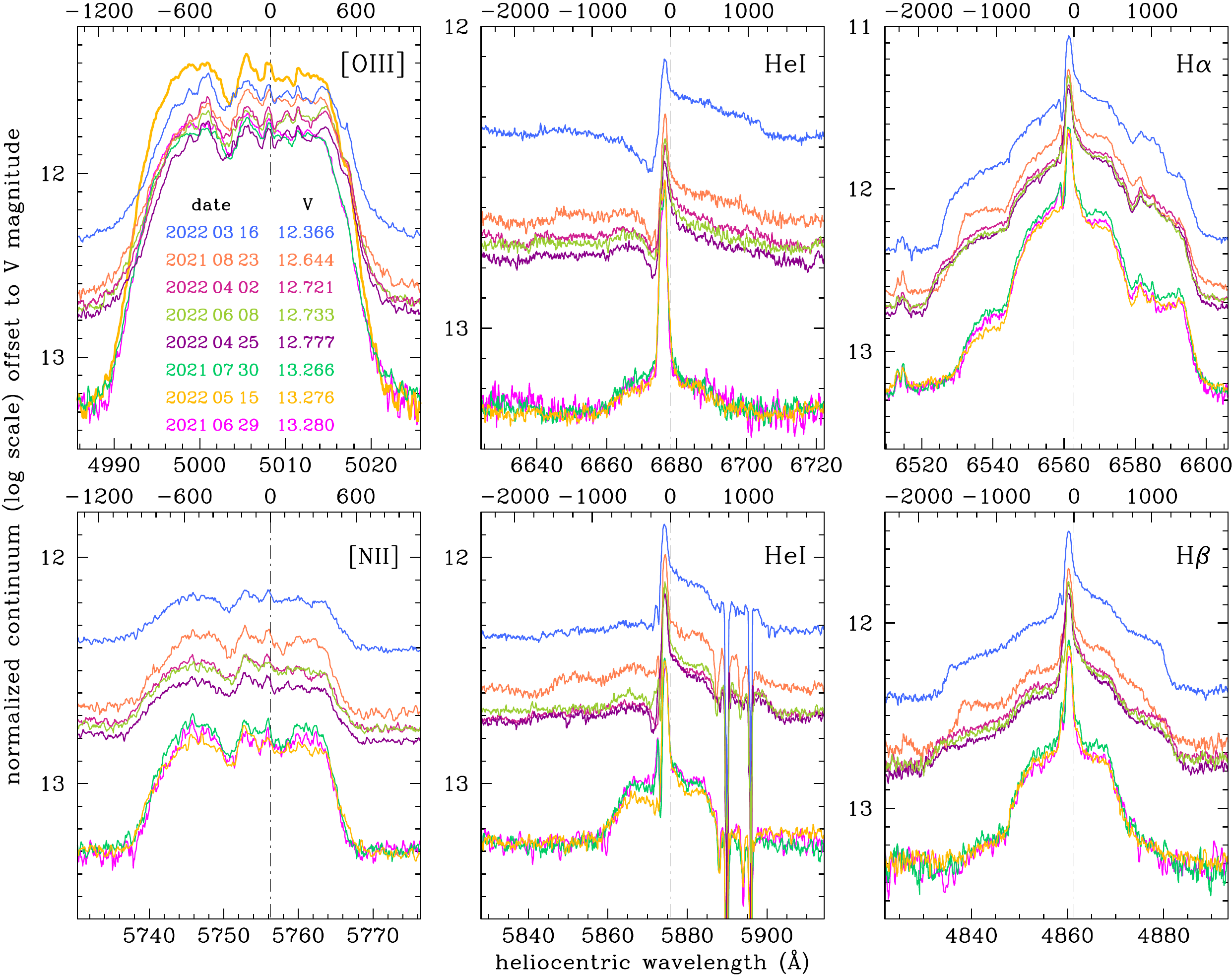}
      \caption{High-resolution profiles of selected emission lines from the
               CHIRON spectra of V5856 Sgr listed in Table~\ref{tab:log}. 
               The logarithm of the continuum-normalized spectrum is plotted
               with an offset equal to the $V$-band magnitude of the nova
               for that date (see legend in  top left panel).  The
               abscissae at the top are velocities in km/s with respect to
               the laboratory wavelength.}
         \label{fig:Chiron}
   \end{figure*}

For the origin of this pseudo-periodicity, an orbital modulation seems
unlikely given its seldom and sudden appearance and the many superimposed
irregularities.  In addition, such a long orbital period would imply an
evolved and consequently bright companion to the WD, which sharply contrasts
with the very large amplitude of the outburst and the nondetection by 2MASS
in quiescence \citep{2003yCat.2246....0C}.  The presence of an evolved
companion also contrasts  with the absence of early nonthermal radio and
X-ray emission \citep{2021ApJS..257...49C, 2021ApJ...910..134G}, which is   instead
regularly observed in novae erupting within symbiotic binaries
\citep[e.g.,][]{2020A&A...638A.130G, 2022MNRAS.514.1557P}, where the
material fed to the circumstellar space by the evolved companion is
violently impacted by the fast nova ejecta.  Some kind of (radial) pulsation
in the envelope could perhaps be a viable explanation; however, V5856~Sgr
lies at a distance from the period--luminosity relation for normal pulsating
stars \citep{2018A&A...619A...8G}, and also its position on the HR is away
from the instability strip.  At the large dimension derived below in Sect. 
5.1 ($\sim$25~R$_\odot$), it is quite possible that the swollen shell of the
burning WD engulfs the companion, and some instability driven by such a
common-envelope arrangement may contribute to the observed 15--17-day
pseudo-periodicity.

\subsection{Deep minimum of 2022 May}

The most prominent event of the 2021-2022 light curve in
Figure~\ref{fig:BVRI} is the {\it deep minimum} (DM) and the subsequent
recovery that V5856 Sgr exhibited around 2022 May.  The photometric colors
changed markedly during the DM, following an intriguing pattern: while
completely flat in $R$$-$$I$, the variations in $B$$-$$V$ were large and
specular to those affecting $V$$-$$R$, clearly indicating that the V band is
the culprit in the observed changes of the colors.  A similar behavior also
characterizes the photometry for 2021 in Figure~\ref{fig:BVRI}, although
with proportionally smaller changes in the colors.

In their analysis of color behavior in novae, \citet[][cf.  Nova Mon 2012
in their Fig.~1]{2013MNRAS.435..771M} noted how the $V$ band  is highly
responsive to the flux emitted in [OIII] 4959+5007, which can account for
$\geq$0.5~mag of the whole flux recorded through the $V$ passband when the
optical spectra are dominated by [OIII], as it is the case for the V5856 Sgr
(see   Figure~\ref{fig:122m}).
The median magnitudes and colors of V5856 Sgr in the months leading up to DM
(February through April 2022) were $V$=12.721, $B$$-$$V$=+0.577,
$V$$-$$R$=+0.439, and $R$$-$$I$=+0.255.  The passage at minimum $V$-band
brightness occurred around May 7.0 UT at $V$=13.508 (see  
Table~\ref{tab:BVRI}), but the extrema in the colors were reached only a
week later around May 14.5 UT at $B$$-$$V$=+0.942 and $V$$-$$R$=+0.074.  The
{\it blueing} of $V$$-$$R$ by 0.365 mag and the {\it reddening} of $B$$-$$V$
by an identical 0.365 mag points to a large strengthening of [OIII]
4959+5007 relative to the underlying continuum.

By a lucky coincidence, we have CHIRON spectra of V5856 Sgr for 2022 April
25 and June 8, which correspond to immediately before and soon after the
changes in colors, and for May 15 when colors deviated the most.  The fluxed
profiles of [OIII] 5007 from these three CHIRON spectra is presented in the
top left panel of Figure~\ref{fig:Chiron}, and confirms the suspected
increase in the flux of [OIII] 4959+5007 at the time of maximum color
change.

   \begin{figure}
   \centering
   \includegraphics[width=\hsize]{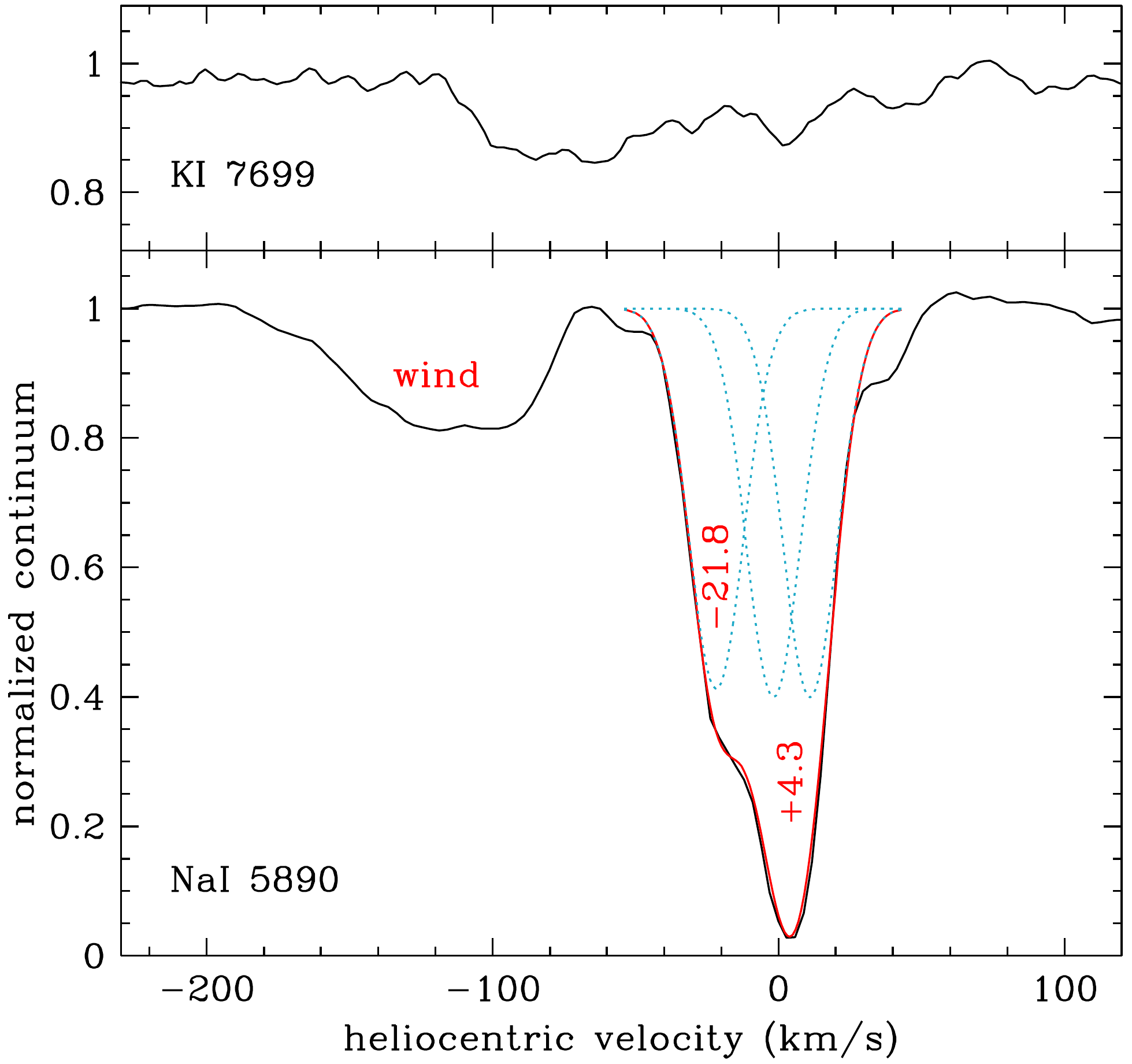}
      \caption{Region around NaI 5890 and KI 7699 from averaged CHIRON
       spectra listed in Table~\ref{tab:log}.  The interstellar components
       at $-$21.8 and $+$4.3 km/s velocity are indicated, as is their
       decovolution (discussed in Sect.  4.3).  The broad
       absorption labeled ``wind'' is variable from spectrum to spectrum.}
         \label{fig:EBV}
   \end{figure}

There may be more behind the DM, however.  Its shape in brightness is not
replicated by the variation of the colors in Figure~\ref{fig:BVRI}, and
the light minimum and color extrema are out of phase by approximately ten days.  The other
minima in Figure~\ref{fig:BVRI}, like those occurring during October and
November 2021, changed   colors proportionally much less than the DM, and
these changes were in the same direction for all colors, not specular as
during the DM.  It seems that the {\it spectroscopic} changes (surge in
intensity of [OIII]) that drove the color changes happened by chance at the
same time of the {\it photometric} DM deep minimum, but the two events may
actually be unrelated.

\section{Reddening}

Of critical relevance to the determination of the radiated luminosity is a
robust knowledge of the interstellar reddening affecting V5856 Sgr, which
appears fairly well constrained to
\begin{equation}
E_{B-V} =0.32 \pm0.03 \label{EBV} 
\end{equation}
by the converging results of the independent methods outlined below.

\subsection{Photometric properties of the main outburst}

Reddening determination from optical photometric colors during the main 2016
outburst of V5856 Sgr were thoroughly investigated by \citet{2017MNRAS.469.4341M}:
the average value from four different photometric criteria is
$E_{B-V}$ = 0.30$\pm$0.05.

\subsection{Three-dimensional dust maps}

V5856 Sgr is located within a few degrees of the direction to the Galactic center,
at a distance of 6.4-7.0~kpc, following \citet{2017MNRAS.469.4341M}, for which 
the 3D reddening maps of \citet{1998ApJ...500..525S} and \citet{2011ApJ...737..103S}
provide $E_{B-V}$ = 0.35 and 0.31, respectively.

\subsection{Interstellar atomic lines}

The CHIRON spectra of V5856 Sgr were examined for the presence of  NaI
5890, 5896 and KI 7699 lines of   interstellar origin.  To increase the
S/N, we  first continuum-normalized all CHIRON spectra listed in Table~2
and then averaged them to produce Figure~\ref{fig:EBV}, which also shows a
broad component in the  NaI lines (highly variable among different spectra, see
the bottom central panel of Figure~\ref{fig:Chiron}), which  we attribute to
a low-velocity wind component, discussed below in Sect. 6.

The interstellar NaI in Figure~\ref{fig:EBV} present at least two clearly
separated components centered at heliocentric velocities $-$21.8 and
$+$4.3~km\,s$^{-1}$, respectively 0.287 and 0.590~\AA\ in equivalent width. 
A Gaussian deconvolution shows the $+$4.3~km\,s$^{-1}$ to be almost two
times broader than the $-$21.8~km\,s$^{-1}$ component (36 vs 22~km\,s$^{-1}$
in FWHM), suggesting it is the blend of two unresolved components of
probably similar intensity, albeit shifted in velocity.  Taking the
$-$21.8~km\,s$^{-1}$ component as the profile of an unblended interstellar
line, we deconvolved the blend at $+$4.3~km\,s$^{-1}$ into two components of
equal intensity.  The three resulting interstellar components are then
plotted as dotted lines in Figure~\ref{fig:EBV}, and their sum is compared
to the observed profile, providing a nearly perfect match.  Their
heliocentric velocities and equivalent widths are $-$21.8, $-$1.8, and
$+$11.0 ~km\,s$^{-1}$, and 0.287, 0.295, and 0.295 \AA, respectively. 
Applying to their equivalent widths the calibration of
\citet{1997A&A...318..269M}, we find the corresponding $E_{B-V}$=0.107,
0.111, and 0.111 reddening values, for a total $E_{B-V}^{tot}$=0.33. 
Quantifying its uncertainty is difficult in view of the arbitrary splitting
of the $+$4.3~km\,s$^{-1}$ blend into two components of equal intensity and
the nonlinear relation between equivalent width and reddening.  Any gross
deviation from equal intensity, however, would have manifested in a
nonsymmetrical profile for their blend, contrary to the observed profile. 
For these reasons we conservatively estimate the error to be $\pm$0.06.

The KI 7699 profile for V5856 Sgr in Figure~\ref{fig:EBV} is rather noisy
and perturbed by the telluric absorptions, which are quite strong in this
spectral region (and wander around when adding spectra in heliocentric
velocity), and probably also by a wind component that seems to be at a lower
velocity than for NaI.  Nonetheless, the KI 7699 profile in
Figure~\ref{fig:EBV} confirms that the +4.3~km\,s$^{-1}$ component seen in
NaI must be the result of the blending of weaker individual components. 
Following the analysis in \citet{1997A&A...318..269M}, if the
+4.3~km\,s$^{-1}$ component is a single line and not the result of a blend,
its large 0.590~\AA\ equivalent width indicates line-core saturation,
resulting in $E_{B-V}$$\geq$0.55.  At such high reddening the equivalent
width of KI 7699 is $\geq$0.14~\AA, and the corresponding line would stand
out clearly in Figure~\ref{fig:EBV}, contrary to evidence.  If the NaI
component at +4.3~km\,s$^{-1}$ is instead the blend of two weaker lines, as
derived above
 the corresponding equivalent width of KI 7699 blend would be only
$\sim$0.06~\AA, a level similar to the noise affecting the spectrum of
Figure~\ref{fig:EBV}.

\subsection{Ultraviolet 2175 \AA\ hump}

The interstellar extinction curve is characterized by the presence of a
broad hump centered around 2175~\AA.  The $UVM2$ photometric
band of the {\it Swift} UVOT telescope includes this hump in its profile,
and it is therefore sensitive to the amount of reddening.  Adopting the
interstellar extinction curve of \citet{1999PASP..111...63F}, to make the
spectral energy distribution of V5856 Sgr  run smoothly through the
three UVOT bands in Figure~\ref{fig:SED} requires dereddening by
$E_{B-V}$=0.30.  A definition of a formal error is not trivial considering
that the details of the interstellar extinction curve may depend on the
given sightline and the loose definition about a {\it smooth} behavior for
the spectral energy distribution of V5856 Sgr through the three UVOT bands. 
A change of $\pm$0.05 to $E_{B-V}$=0.30,  however, would  be perceived by the
eye as breaking such a smooth appearance.

\section{Spectral energy distribution and energetics}

   \begin{figure}
   \centering
   \includegraphics[width=\hsize]{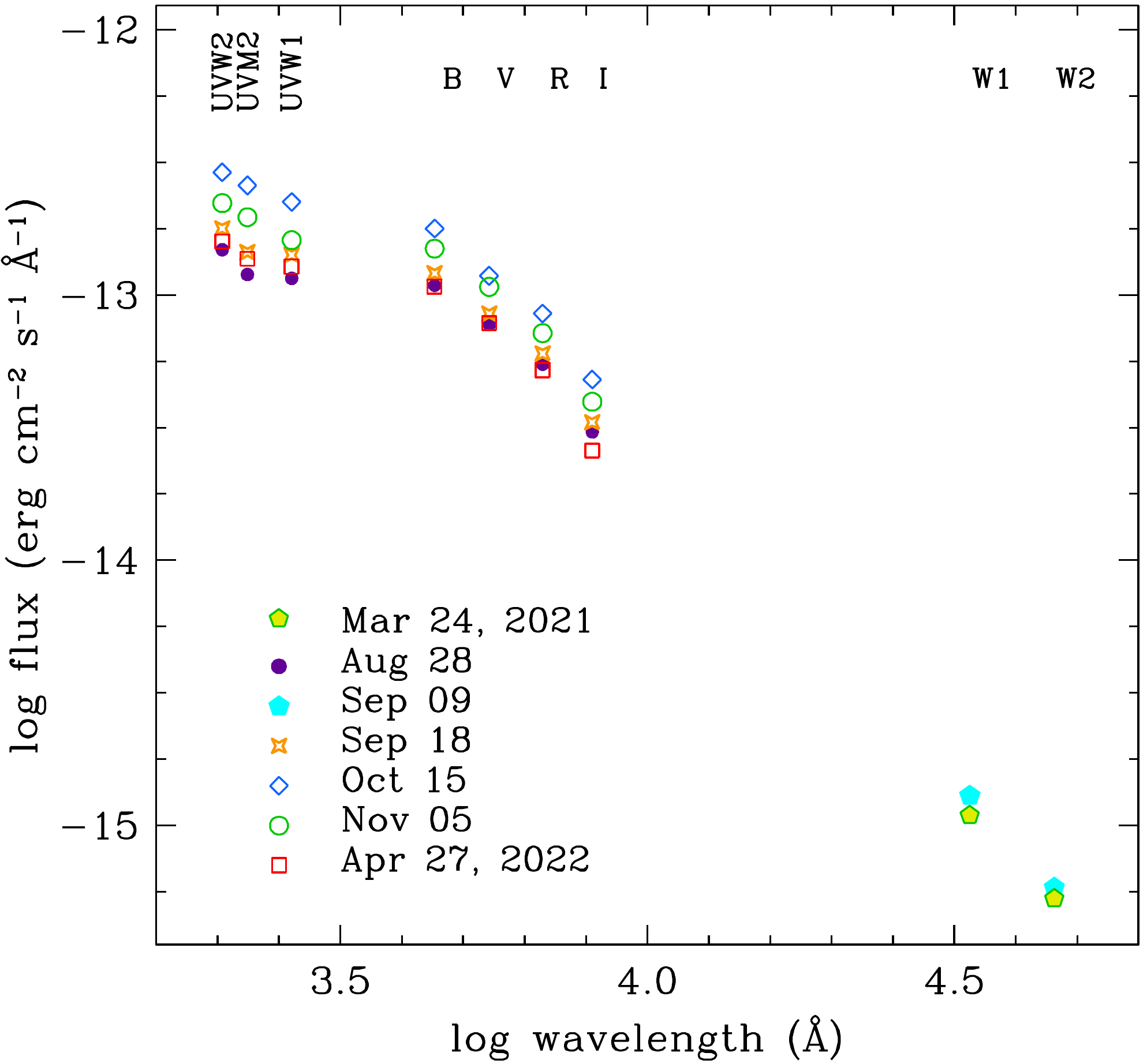}
      \caption{Reddening-corrected spectral energy distribution of V5856
      Sgr at the epochs of the five Swift observations (see Sect.~5 for
      details).}
         \label{fig:SED}
   \end{figure}

   \begin{figure*}
   \centering
   \includegraphics[angle=270,width=13cm]{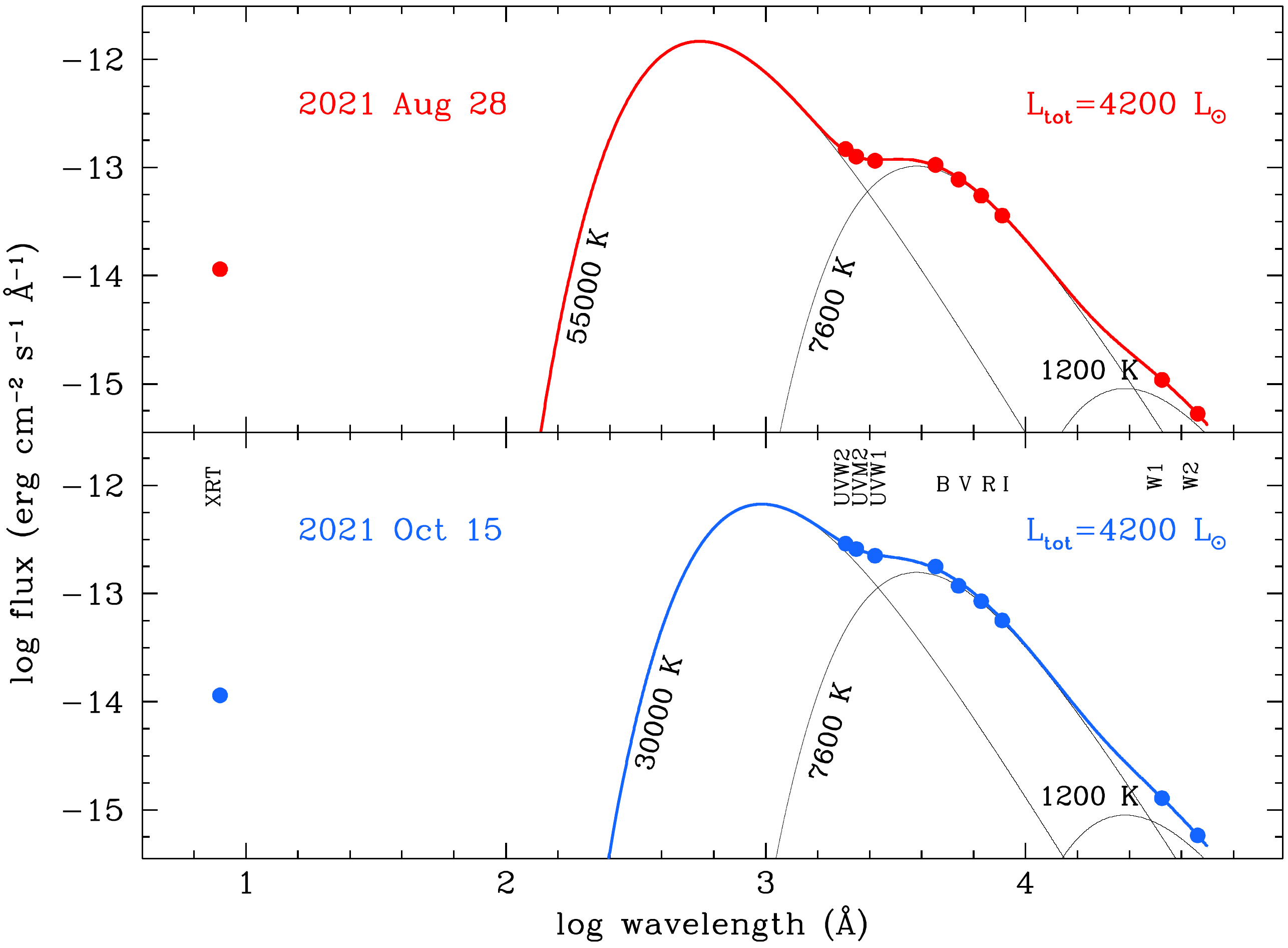}
      \caption{Fit with three blackbodies to the brightest and the
               faintest of the reddening-corrected spectral energy
               distributions in Figure~\ref{fig:SED}.}
         \label{fig:SEDfit}
   \end{figure*}

The spectral energy distributions of V5856 Sgr on the five dates with a {\it
Swift}-UVOT observation is presented in Figure~\ref{fig:SED}.  They are
built combining the ultraviolet fluxes in Table~\ref{tab:Swift} with
$B$$V$$R$$I$ photometry for the same dates from Table~\ref{tab:BVRI} and the
NeoWISE observations of 2021 (centered on March 24 and September 9), and are
reddening-corrected for $E_{B-V}$=0.32.  

\citet{2017MNRAS.469.4341M} estimated a large distance to V5856 Sgr, 6.4-7.0
kpc, from the photometric properties of the outburst; the position on the
sky of the nova ($l$=004.29$^\circ$, $b$=$-$06.46$^\circ$) suggests a
partnership with the Galactic Bulge; the Galactic reddening map of
\citet{2019ApJ...887...93G} returns a lower limit of 5~kpc to V5856 Sgr, and
that of \citet{2019MNRAS.483.4277C} a distance $\sim$8~kpc; finally, the
apparent underluminosity at radio wavelengths \citep{2021ApJS..257...49C}
and the lack of detection in X-rays during the outburst
\citep{2021ApJ...910..134G} both call for a large distance to V5856 Sgr
($>$6 kpc).  For these reasons we assume  for the nova the same distance
as the Galactic Bulge, 8.5 kpc, and scale the resulting energetics to it.

Integrating the flux of the distributions in Figure~\ref{fig:SED} over the
0.16-4.6~$\mu$m interval covered by observations results in the following
luminosities: 
\begin{eqnarray} 
\frac{L_{(0.16-4.6 \mu m)}}{\left( \frac{D}{8.5 ~{\rm kpc}} \right)^2}&=& \label{L016-4.6} \\
&&6.74\times 10^{36}   {\rm erg/s} = 1740 ~L_\odot ~~[2021\,{\rm Aug}\,28] \nonumber\\ 
&& 7.20\times 10^{36} {\rm erg/s} = 1865 ~L_\odot ~~[2021\,{\rm Sep}\,18] \nonumber \\ 
&& 1.14\times 10^{37} {\rm erg/s} = 2950 ~L_\odot ~~[2021\,{\rm Oct}\,15] \nonumber \\ 
&& 8.18\times 10^{36} {\rm erg/s} = 2120 ~L_\odot ~~[2021\,{\rm Nov}\,05] \nonumber \\
&& 7.45\times 10^{36} {\rm erg/s} = 1925 ~L_\odot ~~[2022\,{\rm Apr}\,27] \nonumber \end{eqnarray}
for an average of  $L$=2120~L$_\odot$. These are lower limits to the actual values considering that the maximum
is   located at shorter wavelengths than covered by the  UVOT observations.

To better constrain the actual luminosity radiated by V5856~Sgr, in
Figure~\ref{fig:SEDfit} we  fitted with blackbodies the two
distributions that in Figure~\ref{fig:SED} are characterized by the
brightest (2021 August 28) and the faintest (2021 October 15) fluxes recorded by
UVOT.  A combination of three blackbodies were considered, one for the main
component dominating at optical wavelengths and the  other two to cover the UV
and near-IR excesses.  A combination of blackbodies clearly underrepresents
the true shape of the SED; however, we are only  interested in  an approximate
value for the bolometric luminosity.  A more sophisticated modeling of the
three components would require a much greater  number of parameters than the
only nine photometric values available to sample the SED.

An initial unconstrained run returned a total radiated luminosity of
$\sim$4100~$L_\odot$ for 2021 August 28, and $\sim$4300~$L_\odot$ for 2021 October
15.  Considering that a WD in stable nuclear burning conditions is expected
to radiate at constant luminosity \citep{1971AcA....21..417P}, we 
imposed the condition that the sum of the luminosities of the three fitted
blackbodies is the same at both epochs, and equal to the mean of the above
two unconstrained fits
\begin{equation} 
\frac{L}{\left( \frac{D}{8.5 ~{\rm kpc}} \right)^2} =
1.62\times 10^{37} {\rm erg/s} = 4200 ~L_\odot \label{Ltot} 
,\end{equation}
which corresponds to the hydrogen burning of material of solar composition
at a rate of 
\begin{equation} \dot{M} = 5.7 \times 10^{-8} ~~~{\rm M}_\odot
{\rm yr}^{-1} \label{Mdot} 
.\end{equation}

The total luminosity of the burning shell is thought to be well represented
by the core--mass-luminosity relation of \citet{1971AcA....21..417P}
\begin{equation}
L = 60,000 ~(\frac{M_{\rm WD}}{M_\odot} - 0.522) ~~~~ {\rm L}_\odot \label{Pacz}
,\end{equation}
which returns a mass of 0.6~M$_\odot$ for the WD in V5856 Sgr,  in
good agreement with more recent investigations on steady H-burning at the
surface of a WD by, among others, \citet{2007ApJ...663.1269N},
\citet{2007ApJ...660.1444S}, and \citet{2013ApJ...777..136W}.  A low-mass WD
is also favored by the low expansion velocity of the ejecta observed at the
time of the nova eruption \citep[cf.][and references
therein]{1995cvs..book.....W}.

If $L$=4200~L$_\odot$ has remained constant during the 2018.0--2022.5
plateau, the total amount of radiated energy during this period is
\begin{equation}
E^{(2018.0-2022.5)} = 2.15 \times 10^{45} \left( \frac{D}{8.5 ~{\rm kpc}} \right)^2 
~{\rm erg}  \label{2018-2022}
,\end{equation}
which corresponds to the hydrogen burning of 2.4$\times 10^{-7}$ M$_\odot$
of material of solar composition.  This is a small quantity of material to
be retained by the WD on its surface compared to the amount of mass ejected
in a nova outburst, which is generally estimated in the 10$^{-4}$ to
10$^{-6}$~M$_\odot$ range \citep{1998PASP..110....3G}.

Some comments are now in order about the three components of the SED fitting in
Figure~\ref{fig:SEDfit}.

\subsection{Main component}

The main component to the SED decovolution in Figure~\ref{fig:SEDfit} is a
7600~K blackbody, fitting  both data sets similarly well (within $\pm$200~K). 
Its luminosity is 1325~L$_\odot$ on 2021 August 28 and 2030~L$_\odot$ on 2021
October 15.  The corresponding radii are 21 and 26~R$_\odot$.  The temperature and
dimensions roughly match those of an F0~II/Ib bright giant.  We identify
this main component as the shell of the WD inflated by the stable nuclear
burning at its base.  As we  discuss  in Sect.~6, a constant wind
is blowing off this shell, which the observations indicate produces an effective
photosphere for the visible-IR radiation.

\subsection{Hot component}

The temperature of the hotter component in Figure~\ref{fig:SEDfit} varies
from 3$\times$ to 5.5$\times$10$^{4}$~K, in good agreement with the estimate
from the spectral appearance of V5856~Sgr in Figure~\ref{fig:122m}, following
\citet{1994A&A...282..586M}.  The hot component varies in anti-phase with
the main component at 7600~K, as if some reprocessing may be at play from the hot
to the main component.  The luminosity and radius of the hot component are
2825~L$_\odot$ and 0.6~R$_\odot$ on 2021 August 28, and 2120~L$_\odot$ and
1.7~R$_\odot$ on 2021 October 15.  Given the low temperature of the main
component, the emission lines observed in V5856 Sgr appear powered by the
hot component.  Its location within V5856 Sgr is uncertain, but it could be
related to the polar regions of WD shell, from where  the
faster wind discussed in Sect.~6 below probably originates.

\subsection{Dust}

The same set of NeoWISE observations for 2021 September 9 was included in
the fit to both dates in Figure~\ref{fig:SEDfit} as the closest in time to
the optical and UVOT data.  The resulting 1200 K blackbody probably traces
emission originating from warm circumstellar dust and radiates L$_{\rm
IR}$$\sim$52 L$_\odot$, for a blackbody radius of 165~R$_\odot$, which is
widely external to the main and hot components.  The dust, however, could be
located at greater radii if, instead of being arranged spherically, it
formed  in the equatorial belt discussed   in Sect.~7.

The infrared data available for V5856 Sgr do not allow  the
physical properties of the dust to be constrained.  For the sake of discussion, we can assume that
the dust grains condensing in V5856 Sgr follow the mean properties observed
in other novae \citep{1988ARA&A..26..377G, 1996ApJ...470..577M,
1997MNRAS.292..192E, 1998PASP..110....3G}, meaning that  they are small carbon
grains (radius $a$$\leq$1~$\mu$m, density $\rho$$\sim$2.3~gr~cm$^{-3}$), for
which the Planck mean emission cross section goes as $Q_e = 0.01 a
T^{2}_{dust}$.  Under these assumptions, the mass of the radiating dust in
V5856 Sgr can be estimated as
\begin{equation} M_{dust} = 1.17 \times 10^6 \rho T^{-6}_{dust} \left(
\frac{L_{\rm IR}}{L_\odot} \right) = 3.2\times 10^{-11}~~~{\rm M}_\odot \label{Mdust}
,\end{equation} 
which is rather low and unable to contribute to the reddening affecting V5856 Sgr.

The remarkable stability of the NeoWISE W1, W2 light curve in
Figure~\ref{fig:Overall} suggests that the dust responsible for the infrared
excess cannot be associated with the expanding ejecta of the initial
outburst, bound to cool and fall into oblivion as they disperse into the
surrounding void \citep{1990LNP...369..138G}.  New dust must instead be
forming regularly in the wind constantly blowing off the central star.  The
high $T_{dust}$$\sim$1200~K temperature indicates that the dust grains
condense close to the central star, as close as is allowed by their
sublimation temperature.  As discussed  in Sect.~6, V5856~Sgr
loses mass via winds at different velocities, and the shock interface
where they collide could also be a suitable environment for dust condensation
\citep[e.g.,][]{2017MNRAS.469.1314D}.

\subsection{The X-ray component}

The 3--5.5$\times$10$^{4}$~K temperature of the hot component in
Figure~\ref{fig:SEDfit} cannot account for the X-ray flux recorded by {\it
Swift}, which amounts to $L_X$=0.56~L$_\odot$ over the 0.3-10 keV range for
an unabsorbed bremsstrahlung temperature of kT = 1.0 keV (value averaged
over the five pointings by {\it Swift}).  This X-ray component probably
forms in the outflow from the central star.  Unfortunately, there are not enough
collected X-ray photons to attempt any spectral modeling, for example  to
quantify the contribution of the super-soft emission associated with the
nuclear burning, which is expected to be heavily absorbed from within the WD
inflated shell.

\section{Complex wind outflow}

The gravity at the surface of the $T_{\rm eff}$=7600~K, $R$=21$-$26~R$_\odot$
main component in Figure~\ref{fig:SEDfit} is probably rather low, $\log
g$$\sim$1.4--1.8, depending on the total mass of the WD and its orbiting
companion, surely much lower than $\log g$$\sim$2.5 characterizing a normal
F0~II/Ib bright giant \citep{1981Ap&SS..80..353S}.  From the
\citet{1977A&A....61..217R} expression for mass loss, a wind is expected to
blow off the main component at a rate ($L$, $R$, and $M$ in solar units)
\begin{equation}
\dot{M}_{wind}=4\times 10^{-13}\left( \frac{L\,R}{M} \right)=5 \times 10^{-8} 
~~~{\rm M}_\odot {\rm yr}^{-1} \label{Mwind}
,\end{equation}
similar to the rate at which hydrogen is burnt in the shell of the WD (cf. 
Eq.~(\ref{Mdot})).  Therefore, the mass in the WD shell available to sustain
hydrogen burning reduces at a rate $\approx$1$\times$10$^{-7}$
M$_\odot$\,yr$^{-1}$.

   \begin{figure}
   \centering
   \includegraphics[width=\hsize]{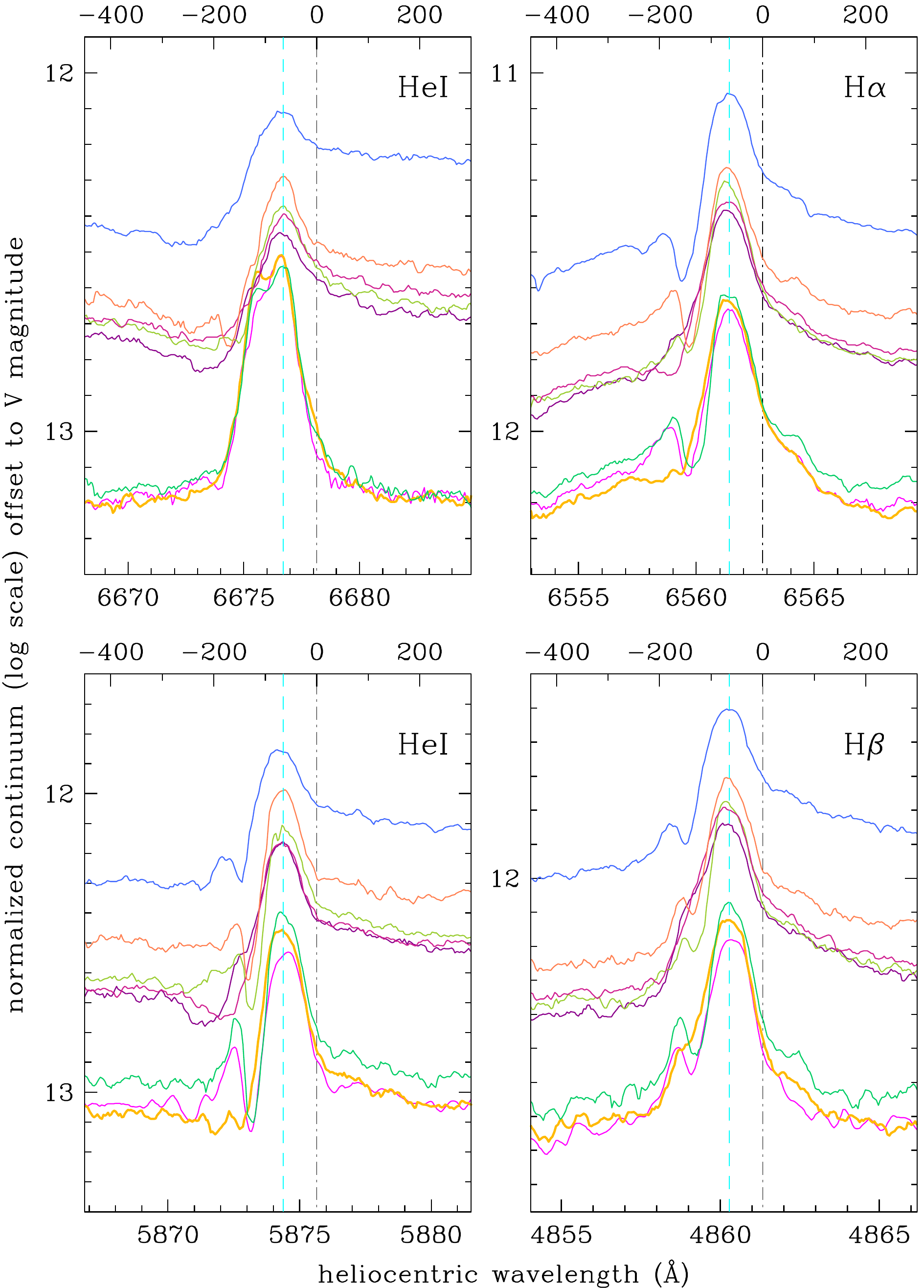}
      \caption{Zoomed-in view from Figure~\ref{fig:Chiron} of the central
      peak displayed by permitted emission lines.  The abscissae at the top
      are velocities in km/s with respect to laboratory wavelength
      (dot-dashed vertical lines).  The dashed vertical line is  the
      $-$65~km/s velocity discussed in Sect.~6.}
         \label{fig:core}
   \end{figure}

The spectra of V5856 Sgr during the plateau indeed provide evidence for a
sustained wind blowing off the central star, as illustrated by the
high-resolution profiles for selected emission lines in
Figure~\ref{fig:Chiron}.  The spectra in this figure are plotted according
to the brightness of V5856~Sgr in the $V$ band, and clearly show how the
profile displayed by the emission lines is dependent on the system
brightness, the dividing line being $V$$\sim$13.0 mag.

When the system is fainter than that (observing epochs {\bf 2021 June 29,
2022 May 15, and 2021 July 30} or the lowest three profiles in each panel of
Figure~\ref{fig:Chiron}), the profiles are dominated by a trapezoidal
pedestal with a full width at zero intensity (FWZI)$\sim$1600 km/s,
similarly present for both permitted and forbidden transitions.  When V5856
Sgr turns brighter than $V$=13.0 mag, a broader and boxier FWZI$\sim$3400
km/s pedestal adds to the profile of permitted lines, but not to that of
forbidden lines.  H$\alpha$ is the strongest permitted line in the optical
spectra of V5856 Sgr, and it shows the FWZI$\sim$3400 km/s component at all
epochs, even if it is weaker when the nova is fainter than $V$=13.0 mag. 
So, it seems appropriate to say that the FWZI$\sim$3400 km/s pedestal {\it
reinforces} when the system is bright, turning visible also for permitted
lines much weaker than H$\alpha$.  The integrated absolute flux of the
FWZI$\sim$1600 km/s pedestal appears to decrease with the system brightness,
the reverse holding true for the FWZI$\sim$3400 km/s component.

The $V$=13.0 mag threshold also drives  another striking difference in
Figure~\ref{fig:Chiron}, this one affecting the  HeI lines.  When V5856 Sgr is
fainter the FWZI$\sim$1600 km/s looks symmetric, for both triplet (5876 Ang)
and singlet (6678) lines, but when the system turns brighter, its blue half
goes missing, and the same also happens  to the FWZI$\sim$3400 km/s
component.  It looks as if  they get chewed up by a wide P~Cyg absorption.  At
the same time, a P~Cyg minimum at about $-$250~km/s appears in HeI 6678,
without much of a counterpart in HeI 5876.

Superimposed on the broad and two-component pedestal, the  permitted lines show
a sharp peak flanked by one or two low-velocity absorption components, as
illustrated by Figure~\ref{fig:core} that zooms in on the core of the same
profiles presented in Figure~\ref{fig:Chiron}.  A similar peak is not
presented by nebular emission lines.  The heliocentric radial velocity and
width of the narrow emission peak are RV$_{\odot}^{em}$=$-$65$\pm$1 km/s and
FWHM$\sim$650 km/s, respectively.  The absorptions to the blue of the
emission peak are rather variable from line to line and epoch to epoch, with
no obvious relation with the $V$=13.0 mag threshold affecting the two
pedestals.  For HeI 5876, H$\beta$ and H$\alpha$ the main absorption is
positioned around $-$150 km/s, a value similar to that displayed by NaI in
Figure~\ref{fig:EBV}.

In addition to the FWZI$\sim$1600 km/s and FWZI$\sim$3400 pedestals and the
FWHM$\sim$650 km/s narrow peak, there is a fourth velocity component in
V5856 Sgr, the one displayed by FeII emission lines, which is characterized
by a narrow and double-peaked profile with a velocity separation $\Delta
v$$\sim$65~km/s, clearly  illustrated by FeII 6516 (multiplet 40) in the
H$\alpha$ panel in Figure~\ref{fig:Chiron}.  The intensity, width, and
velocity separation of the double-peak FeII line profiles do not change in a
noticeable way with the date or the brightness of V5856 Sgr, and therefore
no $V$=13.0 mag threshold apply to them.  The velocity in the $\Delta
v$$\sim$65~km/s component is well below the $v_{esc}$$\sim$125 km/s escape
velocity for the surface of the $\sim$25~R$_\odot$ inflated shell of the WD.

Summing up, during its current plateau V5856 Sgr has been losing a sizeable mount
of mass via wind, which is organized in multiple emission components, neatly
segregated on kinematical grounds: FWZI$\sim$3400, FWZI$\sim$1600 km/s, and
FWHM$\sim$650 km/s.  Wind absorptions are equally present, for example  the
high-velocity mutilations to the HeI profile when the system is
brighter than 13~mag, and the components at $-$250 and $-$150 km/s.

\section{Conclusions}

The observations we  collected and discussed show that V5856 Sgr, six
years past its nova outburst in 2016,  remains bright, having spent the
last 4.5 yrs at an average $I$=12.05 mag, which is only $\Delta I$=6.5 mag
down from the maximum and still $\Delta I$$\geq$10 mag brighter than quiescence. 
During the current plateau, it is radiating a luminosity
$L$$\sim$4200~L$_\odot$, consistent with stable nuclear burning on a WD of
0.6~M$_\odot$.  V5856 Sgr is the source of a strong wind, which is structured
into three distinct and variable emission components with FWZI$\sim$3400,
FWZI$\sim$1600, and FWHM$\sim$650.  The pseudo-photosphere dominating at
optical wavelengths has $T_{\rm eff}$$\sim$7600~K and $R$$\sim$25~R$_\odot$,
widely engulfing the binary system within.  The mass in the WD shell
available to sustain hydrogen burning reduces at a rate
$\approx$1$\times$10$^{-7}$ M$_\odot$\,yr$^{-1}$, while dust at
$T_{dust}$$\sim$1200~K and L$\sim$52~L$_\odot$ keeps forming in the outflow. 
The WD shell can either be remnant material after the nova eruption or
can originate from enhanced mass-transfer from an irradiated swollen
secondary.

Blowing off the wind requires mechanical work. The ratio of the fluxes
radiated in the three components is 10\%:60\%:30\% for FWHM=650, FWZI=1600,
and FWZI=3400 km/s, respectively, as average values for H$\alpha$ over spectra in
Figure~\ref{fig:Chiron}. Applying these proportions to the mass of the wind
in Eq.~(\ref{Mwind}), the corresponding energy going into blowing the wind is
\begin{equation}
E_{kin} = 1.2\times 10^{32} + 6.1\times 10^{33} + 1.4\times 10^{34}=
2\times 10^{34} ~~~~~{\rm erg} \label{kin}
,\end{equation}
which is negligible with respect to the amount of radiated energy in
Eq.~(\ref{Ltot}).

Given these mass ratios, it may be surprising that no counterpart of the
faster FWZI$\sim$3400 km/s wind is visible in the profiles of nebular lines
in Figure~\ref{fig:Chiron}, which are dominated by the FWZI$\sim$1600~km/s
component.  There are some possible reasons for this: ($a$) the
FWZI$\sim$3400 km/s wind may be slowed down by the FWZI$\sim$1600~km/s
material before it reaches the outer radii where the electronic density is low
enough for the formation of nebular lines, and the X-ray emission recorded by
{\it Swift} could originate from these colliding winds
\citep{1997A&A...319..201M, 2013A&A...559A...6L}, and/or ($b$) the
FWZI$\sim$3400 km/s material is ejected only episodically, and it needs the
right time interval to travel to the outer radii where the electron density
drops below the critical value for collisional de-excitation, and thus detecting the
FWZI$\sim$3400 km/s material in the profile of nebular lines could  
be a matter of observing at the right time after such an ejection.

The profiles of permitted emission lines in V5856 Sgr are similar to those in
V2672 Oph (Nova Oph 2009), which were morphokinematically modelled by
\citet{2011MNRAS.410..525M} with three components tracing an equatorial
belt, polar caps for fast bipolar ejection, and a prolate component for the
primary outflow.  By analogy, we infer that the FWZI$\sim$1600 km/s
pedestal, stable over time and shown by all lines, traces the steady wind
blowing off the shell of the WD in a kind of spherically symmetric spatial
arrangement.  The FWZI$\sim$3400 km/s broader components could relate
instead to an episodic bipolar wind, blown off preferentially along the
polar directions, implying that we are looking at V5856 Sgr from an orientation that is closer to 
pole-on   than edge-on.  Finally, the narrow emission peak
traces material laying on the equatorial plane, characterized by   an
electron density that is too high to allow the formation of nebular lines.  The variable and
low-velocity absorption components, blueshifted by about $-$85$\pm$1 km/s
from the narrow emission peak, may form in a gentle mass-loss from the
equatorial belt.

The inferred approximately pole-on orientation of V5856 Sgr seems to add to the
already rich assortment of oddities displayed by this nova.  From the
decline-time versus  the outburst amplitude of novae derived by \citet[][Figure 5.4
in the book]{1995cvs..book.....W}, the amplitude expected for the
log($t_2$)=0.9 decline-time derived by \citet{2017MNRAS.469.4341M} for V5856
Sgr is $\Delta m$=12~mag. The observed $\Delta I$$\geq$16.4~mag is much
larger, surely one of  the largest on record.  Increasing the orbital inclination
toward edge-on conditions (albeit in contrast with the spectral line profiles)
would widen the outburst amplitude up to $\Delta $=15~mag, still appreciably
short of the observed value.

V5856 Sgr is clearly a nova of many peculiarities, which surely deserves 
deeper investigations of its main outburst and a continued monitoring of 
the protracted plateau phase in which it is currently trapped.

\begin{acknowledgements}
We thank the referee for useful comments.  NM acknowledges financial
support through ASI-INAF agreement 2017-14-H.0 (PI: T.  Belloni).  FMW
acknowledges support from NSF grant AST-1614113.  This publication makes use
of data products from the Near-Earth Object Wide-field Infrared Survey
Explorer (NEOWISE), which is a joint project of the Jet Propulsion
Laboratory/California Institute of Technology and the University of Arizona. 
NEOWISE is funded by the National Aeronautics and Space Administration.
\end{acknowledgements}

\bibliographystyle{aa} 
\bibliography{44498corr.bib} 

\end{document}